\documentclass[11pt,a4paper,twocolumn,twoside]{rho-class/rho}
\usepackage[english]{babel}
\usepackage{graphicx}
\usepackage{subcaption} 
\usepackage{pifont}

\title{Cross-Model Humor Preference Modeling with Cards Against Humanity}

\author[{\textasteriskcentered},a]{Victor Winter}
\author[{\textdagger},b]{Farhan Lakhany}

\affil[a]{University of Nebraska at Omaha}
\affil[b]{San Jose State University}

\corres{\textsuperscript{\textasteriskcentered}Corresponding author: \href{mailto:vwinter@unomaha.edu}{vwinter@unomaha.edu} (University of Nebraska at Omaha).}

\theday{\today}

\begin{abstract}
This paper investigates whether one large language model can approximate the humor preferences of another in a controlled Cards Against Humanity-style task. Two models — GPT-4o as Czar and Claude Opus-4.5 as Player — are evaluated on a binary humor-selection task constructed so that success cannot follow from self-preference. A reflected-cell stability procedure isolates 244 hands on which the two models hold deterministic but opposite preferences, partitioned into a 97-hand context pool and a 147-hand held-out test pool. The Player is then evaluated across five graded conditions: default self-preference, generic Czar-modeling instruction, model-identified Czar, prior Czar selections, and prior Czar selections with rationales. This gradient is designed to separate two sources of improvement: framing effects, in which the Player is told to attend to a Czar without seeing any of the Czar's behavior, and direct behavioral evidence, in which the Player is shown the Czar's prior choices. Player accuracy increased from 0.7\% in Condition 1 to 19.0\% and 25.9\% in the framing-only conditions, and then rose to 72.8\% and 82.3\% once behavioral evidence and rationales were provided. An omnibus Cochran's Q test and pairwise McNemar tests confirmed that each step in the gradient produced a significant improvement. The results indicate that role instruction and model identity yield only modest gains, while behavioral evidence — especially when accompanied by rationales — supports substantial cross-model preference modeling. The findings are interpreted as theory-of-mind-like behavior in an operational rather than representational sense: the Player shifts away from self-preference toward another agent's demonstrated preferences, without any claim about an underlying representation of mental states.
\end{abstract}


\begin{document}

    \maketitle
    \thispagestyle{firststyle}

\section{Introduction}

Humor is a difficult domain for computational analysis because it is subjective, context-sensitive, and socially situated. What one person finds funny may leave another unmoved, and successful humor often depends less on universal rules than on audience-specific expectations. The game \emph{Cards Against Humanity} captures this structure in a simple form: players do not merely choose the card they personally find funniest; they try to choose the card that a particular judge, or Czar, is most likely to select. 

This paper uses a controlled Cards-against-Humanity-style task to investigate whether one large language model can approximate the humor preferences of another. The experiment replaces ordinary gameplay with a structured binary preference-selection task in which one model serves as the Czar and another serves as the Player. The central question is whether the Player can move beyond its own default preference and select the white card preferred by the Czar.

The experiment focuses on a deliberately difficult subset of hands: cases in which GPT-4o and Claude Opus-4.5 exhibit stable but opposite preferences. This opposed-preference design makes the task especially informative. If the Player simply chooses according to its own preference, it should fail. Success therefore requires some form of Czar-preference modeling: predicting another model's evaluative pattern rather than expressing one's own.

The experiment evaluates Player performance under five increasingly informative conditions, ranging from default self-preference to one in which the Player is given examples of the Czar's prior choices and rationales. The results provide a controlled behavioral test of whether an LLM can use role instructions, model identity, examples, and explanations to approximate another model's preference structure in a subjective humor-selection domain. Unlike most existing benchmarks for theory of mind in large language models, which test the tracking of beliefs and goals about states of the world, this experiment focuses on the modeling of another agent's evaluative preferences. Correctness in this content domain is constituted by the evaluator rather than fixed independently of them. Tracking what someone believes about a hidden object is an epistemic task with a fact-of-the-matter answer; modeling what someone finds funny is an evaluative task whose answer depends on the agent being modeled. Where the marble is does not depend on Sally; what is funny to GPT-4o does.

\section{Related Work}

The framing introduced above sits at the intersection of three lines of research: computational humor, large language model preference modeling, and theory-of-mind-like reasoning. Each contributes context for this experiment, but none has directly addressed cross-model humor-preference modeling.

\subsection{Computational Humor and Cards Against Humanity}

Recent work on humor and language models continues to emphasize that humor generation, explanation, and evaluation remain challenging, particularly when humor must be judged in socially situated or subjective contexts rather than as a simple text classification problem. Loakman et al., for example, frame humor generation and explanation as still underdeveloped areas for LLMs, despite rapid progress in other forms of language generation \cite{loakman2025whoslaughingnowoverview}.

The game \emph{Cards Against Humanity} provides a useful experimental domain for computational humor because it reduces humorous response generation to a structured selection problem: given a prompt card and a set of response cards, choose the response that is funniest or most appropriate. Ofer and Shahaf's \emph{Cards Against AI: Predicting Humor in a Fill-in-the-blank Party Game} uses a large CAH dataset to predict winning cards in human gameplay. Their work treats CAH as a humor-prediction problem grounded in historical human choices, whereas the present experiment uses a controlled CAH-style task to study whether one LLM can model another LLM's preference structure \cite{ofer2022cardsaipredictinghumor}.

A more recent and directly related study, \emph{Cards Against LLMs: Benchmarking Humor Alignment in Large Language Models}, evaluates frontier models on CAH-style games using human gameplay as a comparison point. Fettach et al. report that LLMs perform above random baseline but align only modestly with human preferences. They also find that models agree with one another more than they agree with humans, and that model choices are shaped partly by position biases and content preferences \cite{fettach2026cardsllmsbenchmarkinghumor}. This experiment addresses these methodological concerns explicitly through reflected-cell analysis, stability filtering, and randomized white-card order in the final test set.

This experiment differs from these CAH-based studies in its target. Prior CAH work has primarily asked whether models can predict human humor judgments or whether model humor aligns with human preferences. This experiment instead asks whether one LLM can predict the stable humor preferences of another LLM, shifting the focus from human-alignment evaluation to cross-model preference modeling.

\subsection{Theory of Mind and Audience Modeling in LLMs}

The experiment also connects to recent work on Theory of Mind in large language models. Theory of Mind is commonly understood as the ability to attribute mental states to others and use those attributed states to predict behavior. Chen et al.'s 2025 survey \cite{chen-etal-2025-theory} reviews recent efforts to evaluate and improve ToM-like abilities in LLMs, with particular attention to benchmark design and enhancement methods. Nguyen's 2025 survey \cite{nguyen2025surveytheorymindlarge} similarly distinguishes behavioral and representational accounts of ToM in LLMs, emphasizing the need for caution when interpreting successful benchmark performance as evidence of human-like mental-state reasoning.


The Chen et al. survey organizes existing benchmarks against the Abilities in Theory of Mind Space (ATOMS) framework developed by \cite{beaudoin2020systematic}, a taxonomy derived from systematic review of 220 measures across 830 studies. ATOMS identifies seven categories of mental states and social situations: beliefs, intentions, desires, emotions, knowledge, percepts, and mentalistic understanding of non-literal communication. Most current benchmarks cluster around belief tracking, with newer work extending into intentions, emotions, and social goals. OpenToM, for example, builds on the Sally-Anne paradigm by adding character personalities and emotion-related questions, while MuMA-ToM moves into multimodal, multi-agent settings involving belief and goal inference about household activities. FANTOM tests belief tracking in conversational settings designed to expose inconsistencies in surface performance. The shared structure of these benchmarks is that the mental states being attributed concern situations external to the agent: where an object is located, what another character knows, what goal an agent is pursuing in a scene.

Evaluative preferences — taste, humor, aesthetic judgment — are represented in ATOMS \cite{beaudoin2020systematic} only in narrow recognition forms, such as recognizing that another agent has a different desire (the discrepant desires sub-ability) or that a statement is meant as a joke (the humor sub-ability under mentalistic understanding of non-literal communication). ATOMS does not include a sub-ability corresponding to modeling an agent's evaluative preference structure across novel cases, which is the capacity the present experiment probes behaviorally.

The experiment described in this article does not require the Player model to represent the Czar as having beliefs, desires, or subjective awareness. Instead, it asks whether the Player can behaviorally shift from self-preference to other-preference prediction. This shift is operationalized as Czar-preference modeling: selecting the card predicted to match the Czar's demonstrated preference rather than the Player's own default choice.

Recent critiques of LLM ToM benchmarks are also relevant. Riemer et al. \cite{riemer2025positiontheorymindbenchmarks} argue that many ToM benchmarks for LLMs are limited because they often test static social-reasoning patterns rather than adaptation to a new interaction partner. This experiment addresses this concern in a narrow but concrete way: the Player is evaluated on whether it can adapt to a particular Czar's observed preferences, especially when given prior examples and rationales. This makes the task less a test of abstract social reasoning and more a test of partner-specific preference prediction.

\subsection{Preference Modeling and Personalized Alignment}

The experiment also relates to work on preference modeling and personalized alignment, which aims to adapt LLM behavior to particular users or groups rather than assuming a single universal preference target. Xie et al.'s survey \cite{xie2025surveypersonalizedpluralisticpreference} describes this as an emerging research direction and organizes existing methods into training-time, inference-time, and user-modeling approaches.

Several recent systems explore ways to infer or represent user preferences from prior behavior. USER-LLM, for example, contextualizes LLMs using user embeddings derived from user interaction histories, allowing model outputs to adapt to user-specific behavioral patterns \cite{liu2024userllm}. This line of work is relevant because the experiment similarly treats preferences as something that can be inferred from observed choices. The target of modeling, however, differs: personalized alignment typically asks how a model can adapt to a human user, whereas this experiment asks whether one LLM can infer the preference structure of another LLM whose stable preferences have first been identified through a controlled filtering procedure.

\subsection{In-Context Learning from Examples and Rationales}

The largest performance gains in the experiment occur when the Player receives direct evidence of the Czar's prior choices. This connects to in-context learning, where models use examples in the prompt to infer task structure or adapt behavior at inference time.

The additional improvement from including rationales connects to explanation-augmented in-context learning. Krishna et al.'s AMPLIFY framework shows that post hoc explanations can improve model performance when incorporated into in-context learning \cite{NEURIPS2023_ce65173b}. Honda and Oka similarly argue that explanation-based in-context learning can improve robustness, particularly when models must generalize beyond the distribution of the provided demonstrations \cite{honda-oka-2025-exploring}. These findings provide a useful lens for interpreting the difference between Conditions 4 and 5 in this experiment, where rationales are added to the same set of prior Czar selections.

\subsection{Position Bias and Reliability in LLM Judgment}

Finally, the experiment relates to studies of bias and reliability in LLM judgments. LLMs used as evaluators may exhibit systematic artifacts, including sensitivity to presentation order, verbosity, surface features, and prompt wording. In CAH-style tasks, position bias is especially important because the model is asked to choose between candidate responses. Fettach et al. report that LLM humor judgments in CAH-style games are partly explained by systematic position biases and content preferences \cite{fettach2026cardsllmsbenchmarkinghumor}.

The experiment described in this article incorporates this concern into its data model. Each pair of white cards is evaluated in both orientations, allowing the procedure to distinguish stable content-based preference from positional bias (i.e., orientation-sensitive choices). Hands that fail stability or reflected-cell checks are excluded from the final opposed-preference set, providing a cleaner basis for evaluating Czar-preference modeling.

\section{Experiment}

The experiment uses the family edition of \emph{Cards Against Humanity}, which avoids the adult content of the original game while preserving the basic structure of black-card prompts and white-card responses. This makes the materials more suitable for academic analysis and reproducible evaluation.  The Czar's pattern of selections across hands is referred to informally as the Czar's humor and operationally as the Czar's preference structure: humor is treated as a subjective preference domain rather than something reducible to a precise algorithmic definition.

A further methodological benefit is that family-edition content largely sidesteps the safety guardrails of contemporary LLMs. Explicit or transgressive prompts can trigger refusals, hedged completions, or modified outputs, and these behaviors vary considerably across providers and model versions. Such variation would confound the preference signal of interest, conflating humor judgment with provider-specific safety policy. The family edition lets the Czar and Player engage with every hand on the same terms.

\subsection{Materials}

The card corpus consists of:

\begin{center}
\fbox{
\begin{tabular}{ll}
$11$ black cards &   modeled by the array: $black[0..10]$ \\
$32$ white cards &  modeled by the array: $white[0..31]$ \\
\end{tabular}
}
\end{center}

Each black card functions as a prompt containing a blank or implied response location. Each white card functions as a possible completion. For example:

\medskip

\noindent
\fbox{
\begin{minipage}{0.45\textwidth}
You are playing a card game similar to Cards Against Humanity.

\medskip
Here is the prompt card: 

\smallskip

"This is gonna be the best sleepover ever. Once Mom goes to bed, it's time for $\underline{\hspace{10mm}}$!"

\medskip

Here are two possible response cards:

\medskip
Option A: "Kissing Mom on the lips." \\
Option B: "Playing trumpet for the Mayor." 

\medskip
Which option is the funnier, more entertaining response to the prompt card? \\

First, explain your reasoning in 2-3 sentences. Then, state your final selection. \\
\end{minipage}
}

\medskip
This illustrates the basic unit of evaluation: a single black card paired with two possible white-card responses.

\subsection{The CaH-Hand}

The atomic data object in the experiment is the \emph{CaH-hand}: a triple \emph{(black-card, white-card-A, white-card-B)} in which the black card is drawn from the black-card array and two distinct white cards are drawn from the white card array. Presentation order matters because language models may exhibit positional preference. Formally, for black-card index $k$ and distinct white-card indices $i \neq j$:

\[
H(k, i, j) = (black[k], white[i], white[j])
\]

The same underlying pair of white cards therefore produces two ordered hands, $H(k, i, j)$ and its reflected counterpart $H(k, j, i)$.

\subsection{Grid Representation of CaH-Hands}

For each black card $k$, the experiment constructs a $32 \times 32$ grid, denoted $grid[k]$, whose axes correspond to indices in the white-card array. Each cell $cell(i, j)$ denotes the ordered hand $(black[k], white[i], white[j])$. Cells on the main diagonal are excluded because they pair a white card with itself. Each grid therefore contains $32 \times 31 = 992$ valid ordered hands, and the full grid space across all $11$ black cards contains $11 \times 992 = 10912$ valid ordered hands.  

\subsection{Data Model Summary}

\begin{figure*}
\centering
\fbox{
\begin{tabular}{lll}
$black[0..10]$      & = & array of 11 black cards \\
$white[0..31]$     & = & array of 32 white cards \\

\multicolumn{3}{c}{\hspace{5mm}} \\

$grid[k]$           & = & $32 \times 32$ grid associated with $black[k]$ \\

$cell(i,j)$         & = & ordered candidate hand:  $(black[k], white[i], white[j])$ \\

\multicolumn{3}{c}{\hspace{5mm}} \\

valid cell        & = & $cell(i,j)$ where $i \neq j$ \\

reflected cell   & = & $cell(j,i)$ \\

\multicolumn{3}{c}{\hspace{5mm}} \\

stable hand      & = & 
\begin{minipage}[t]{5in}
a hand for which $cell(i,j)$ and $cell(j,i)$ imply the same preferred white card 
\medskip
\end{minipage}
\\

orientation-sensitive hand & = & 
\begin{minipage}[t]{5in}
a hand for which $cell(i,j)$ and $cell(j,i)$ do not imply the same preferred white card, indicating that presentation order may be influencing the model's selection
\medskip
\end{minipage}
\\

deterministic hand & = & 
\begin{minipage}[t]{5in}
a stable hand for which repeated model evaluations yield the same preferred white card 
\medskip
\end{minipage}
\\

borderline hand   & = &
\begin{minipage}[t]{5in} 
a hand for which repeated evaluations by the same model do not yield the same preferred white card, suggesting that neither white card is strongly preferred
\medskip
\end{minipage}
\\

opposed-preference hand & = & 
\begin{minipage}[t]{5in}
a deterministic hand for which two models prefer opposite white cards 
\medskip
\end{minipage}
\\
\end{tabular}
}

\caption{Data Model and Concepts}\label{fig:data-model}
\end{figure*}

Figure \ref{fig:data-model} summarizes the data elements and concepts used in our experiment.

Note that reflected-cell analysis distinguishes content-based preference from positional preference, while repeated evaluation identifies hands without a stable preference. Both filters are needed before two models' grids can be compared meaningfully.

\section{Methods}

The experiment proceeds in three broad phases. First, model-specific preference grids are constructed for each black card. Second, corresponding grids from two models are compared to identify hands for which the models exhibit stable opposed preferences. Third, the resulting opposed-preference hand set is used to evaluate whether a Player model can select the white card preferred by a Czar model under increasingly informative experimental conditions.

\subsection{Model-Specific Grid Evaluation}

\begin{figure*}[htbp!]
    \centering
    \begin{subfigure}[t]{0.49\textwidth}
        \includegraphics[width=\linewidth]{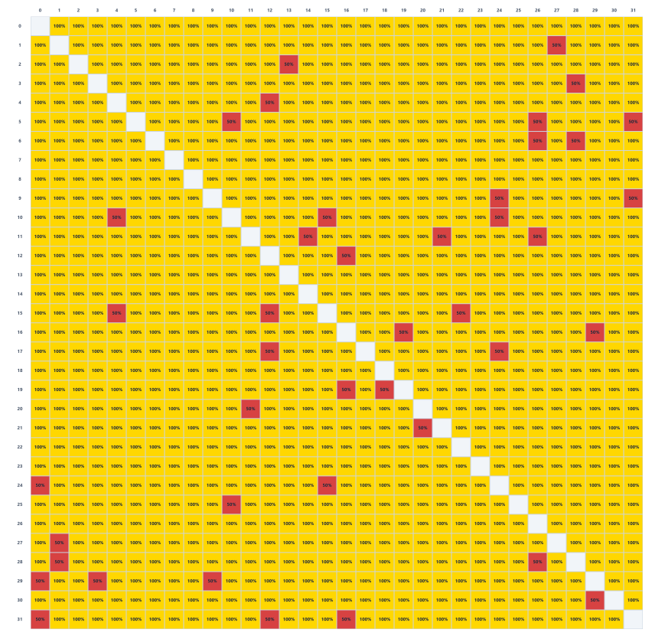}
        \caption{Claude Opus 4.5: Initial Model-Specific Evaluation}\label{fig:model-specific-evaluation}
    \end{subfigure}
    \hfill
    \begin{subfigure}[t]{0.49\textwidth}
        \includegraphics[width=\linewidth]{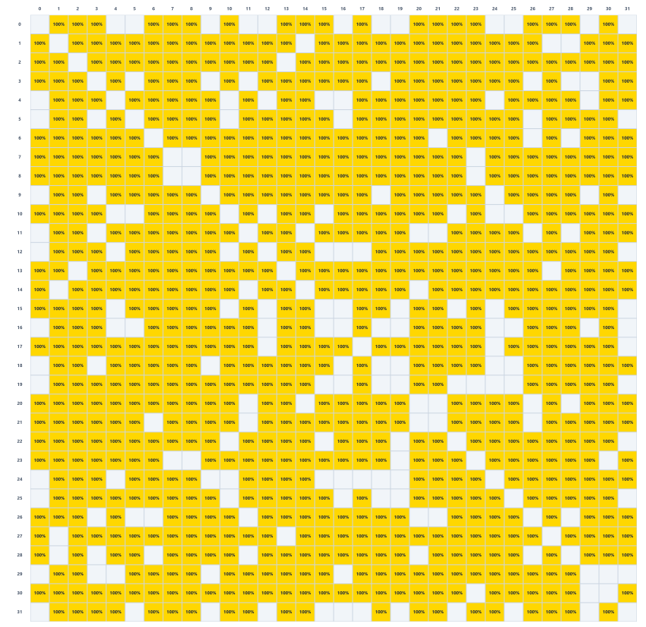}
        \caption{Claude Opus 4.5: After Filtering Disagreement and Positional Bias}\label{fig:filtered-model-specific-evaluation}
    \end{subfigure}

    \begin{subfigure}[b]{0.49\textwidth}
        \includegraphics[width=\linewidth]{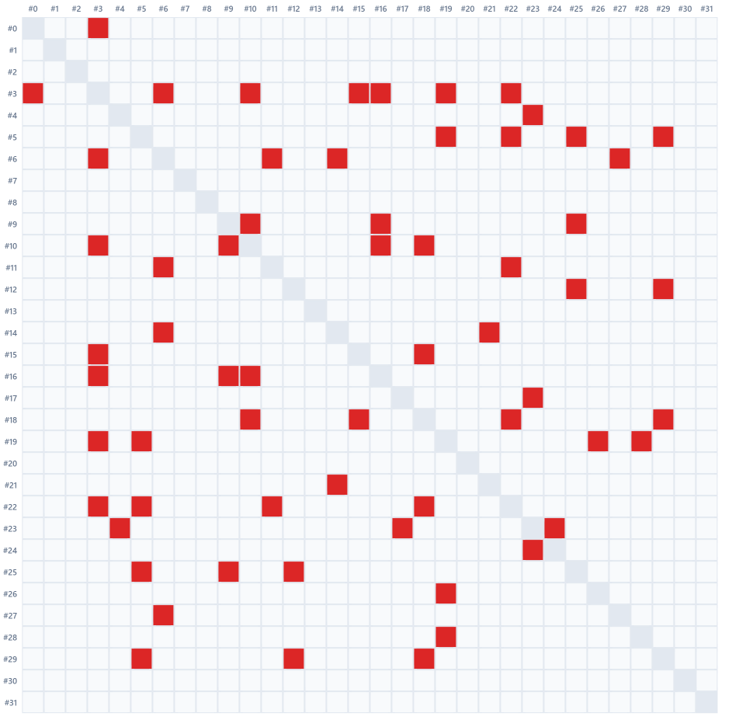}
        \caption{GPT-4o and Opus 4.5: Initial Contention Grid}\label{fig:initial-contention-grid}
    \end{subfigure}
    \hfill
    \begin{subfigure}[b]{0.49\textwidth}
        \includegraphics[width=\linewidth]{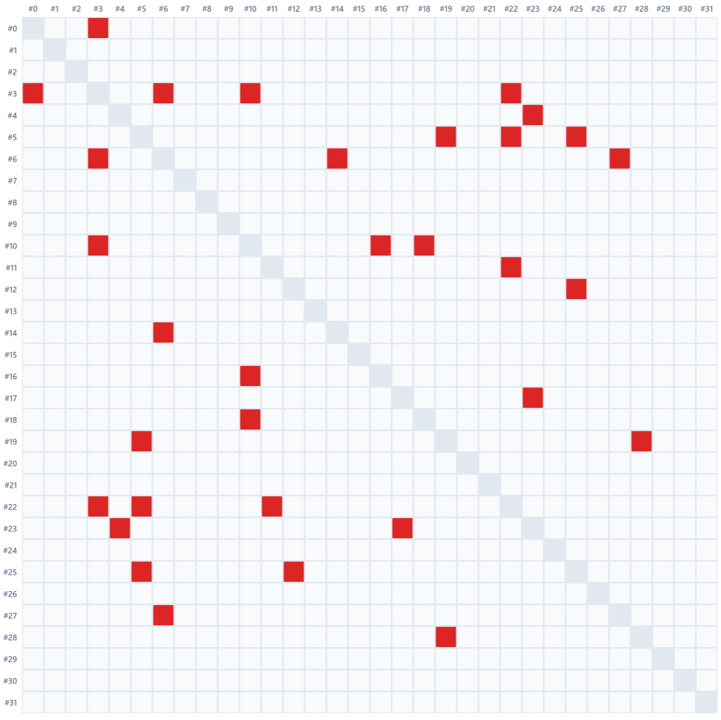}
        \caption{GPT-4o and Opus 4.5: Stable Contention Grid}\label{fig:stable-contention-grid}
    \end{subfigure}

    \caption{Grids associated with black[0]}\label{fig:grids}
\end{figure*}

For each black card, a model-specific preference grid is constructed over the valid ordered hands defined above. For each hand, a model is asked to select the funnier or more entertaining white-card response. In this phase, two separate instances of the same model are used: one as Czar and one as Player. This produces an initial estimate of the model's preference for each valid ordered hand. The procedure is performed independently for each model, yielding \emph{GPT-4o\_grid[k]} and \emph{Opus-4.5\_grid[k]} for each black card $k$. 

Figure \ref{fig:model-specific-evaluation} shows the $grid[0]$ for Claude Opus 4.5.  Figure \ref{fig:filtered-model-specific-evaluation} shows the cells of $grid[0]$ for (Claude) Opus 4.5 that remain when disagreement cells and positional bias cells have been removed.

\subsection{Construction of Contention Grids}

For each black card $k$, the GPT-4o grid and Opus-4.5 grid are intersected to identify hands for which the two models prefer opposite white cards. The resulting filtered grid is called a \emph{contention grid}. For example, the contention grid associated with the card $black[k]$ is denoted $contention\_grid[k]$. A hand is included only if GPT-4o and Opus-4.5 select different white cards for the same underlying hand: if GPT-4o prefers $white[i]$ and Opus-4.5 prefers $white[j]$, or vice versa.

This initial disagreement is only an approximation. Some apparent disagreements may arise from borderline hands in which one or both models do not have a stable preference, so the contention grid must be subjected to additional filtering. Figure \ref{fig:initial-contention-grid} shows the initial state of $contention\_grid[0]$.

\subsection{Stability Filtering and Fixed-Point Re-Evaluation}

When two white cards are approximately equal in humorous value for a given model, repeated evaluations may yield different selections. Such hands are not suitable for the opposed-preference set because they do not represent reliable preference differences.

Each contention grid is therefore subjected to repeated re-evaluation. In each round, all remaining candidate hands are re-evaluated by the relevant models, and a hand is retained only if both models continue to exhibit the same opposed preference observed in the previous round. The process is repeated until a fixed point is reached — that is, until a complete re-evaluation removes no additional hands:

\begin{enumerate}
\item Construct initial $contention\_grid[k]$ (see Figure \ref{fig:initial-contention-grid}).
\item Re-evaluate every remaining hand (i.e., cell) in $contention\_grid[k]$.
\item Remove any hand whose opposed-preference relation fails to persist.
\item Repeat steps 2–3 until no additional hands are removed.
\item Treat the remaining hands as the stable contention grid for $black[k]$.
\end{enumerate}

This procedure is intentionally conservative: it removes hands that are sensitive to stochastic variation, ambiguous preference, or borderline comedic value. The resulting fixed-point grids contain only hands for which the two models have deterministic but opposite preferences. Figure \ref{fig:stable-contention-grid} shows the stable hands (i.e., cells) resulting from the fixed point evaluation of $contention\_grid[0]$.

\subsection{Construction of the Experimental Hand Set}

The $11$ fixed-point contention grids are combined into a single sequence of stable opposed-preference hands. Each hand in this sequence satisfies three requirements:

\begin{enumerate}
\item The hand contains one black card and two distinct white cards.
\item Each model exhibits a stable preference for one of the two white cards.
\item The two models prefer opposite white cards.
\end{enumerate}

This construction creates the central experimental condition: the Player's default preference is known to conflict with the Czar's preference, so success cannot be explained by the Player simply choosing the card it would independently prefer.

\subsection{Context Pool and Test Pool Partitioning}

The stable opposed-preference hand set is partitioned into a \emph{context pool} and a \emph{test pool}. The context pool contains hands that may be shown to the Player as examples of the Czar's prior behavior. The test pool contains held-out hands used to evaluate Player performance. The purpose of this partition is to create conditions where the Player model (Opus 4.5) can be given various opportunities to learn or adapt to the preferences of the Czar model (GPT-4o).

\subsection{Experimental Conditions}\label{section:experimental-conditions}

The experiment evaluates Player performance under five increasingly informative conditions. Each condition presents the same binary choice task, but varies the information available to the Player.

\paragraph{Condition 1: Default Player Preference.} The Player is asked to select the white card it finds funnier or more entertaining. No information about the Czar is provided. This condition serves as a baseline: under the opposed-preference design, the Player's default preference is expected to conflict with the Czar's preference, and the success rate should approach 0\%.

\paragraph{Condition 2: Generic Czar-Modeling Instruction.} The Player is instructed to choose the card that the judge, or Czar, is most likely to prefer, but is not told which model is serving as the Czar. This tests whether a generic shift from self-preference to other-preference reasoning improves performance.

\paragraph{Condition 3: Model-Identified Czar.} The Player (Opus-4.5) is told that GPT-4o is serving as the Czar. This tests whether the Player has an implicit model of GPT-4o's preference tendencies — its style, reasoning patterns, or response habits.

\paragraph{Condition 4: Past Czar Evaluations Provided.} The Player is given examples of past evaluations made by the Czar, drawn from the context pool. Each example contains a previous CaH-hand and the white card selected by GPT-4o. Evaluation is performed on the held-out test pool, so success requires generalizing from prior Czar decisions to new hands.

\paragraph{Condition 5: Past Czar Evaluations plus Rationales.} Each context example now also includes the Czar's rationale. That is, each example in the context pool contains the following information:

\begin{itemize}
\item the identity of the Czar (in our experiment GPT-4o)
\item the black card 
\item the two white-card options 
\item the Czar's selected white card 
\item the Czar's explanation or rationale 
\end{itemize}

A rationale may reveal features of the Czar's preference structure that are not obvious from the selection by itself — for example, whether the Czar favored absurdity, literalness, surprise, social incongruity, escalation, or specificity.

\subsection{Decomposition of the Condition Gradient}

The five conditions are organized into two tiers reflecting qualitatively different sources of potential improvement. Conditions 1–3 form a framing tier: the Player's instructions and information about Czar identity are varied, but the Player is not shown any of the Czar's actual behavior. Conditions 4–5 form a behavioral-evidence tier: the Player is given concrete examples of the Czar's prior choices, with rationales added in Condition 5. The boundary between Condition 3 and Condition 4 therefore separates gains attributable to instruction and model identity from gains attributable to observed Czar behavior. This decomposition is central to the interpretation of results: shifts within the framing tier reflect what role information and general model identity contribute on their own, while shifts at the framing-to-behavioral-evidence boundary and within the behavioral-evidence tier reflect what direct evidence of prior Czar decisions and reasoning adds beyond framing.

\subsection{Black-Card Coverage and Comedy Footprints}

The context pool is constructed with attention to black-card coverage. Experimental observations suggest that CaH humor is not one-dimensional: different black cards create different comedic situations, and features that make a white card successful for one prompt may not transfer cleanly to another. Each black card can be understood as having a \emph{comedy footprint}: a local structure of comedic possibilities shaped by the prompt, the implied blank, and the kinds of responses that can plausibly complete it. While these footprints may overlap, they are not interchangeable. Consequently, the context pool is constructed to include examples associated with every black card represented in the test pool, giving the Player evidence about how the Czar's general tendencies appear within each local comedic structure.

\subsection{Outcome Measure}

The primary outcome measure is the Player's success rate in selecting the white card preferred by the Czar:

\[
\emph{success rate} = \frac{\emph{successful Player selections}}{\emph{total test hands}}
\]

\medskip
For each test hand, the Player's selected white card is compared against the Czar's stable preference as established during the fixed-point contention-grid construction. The five conditions form a graded sequence in the kind of information made available to the Player, ranging from no Czar information at all in Condition 1 to full prior selections plus rationales in Condition 5.

\section{Results}

The experimental hand set was constructed to create a difficult Czar-preference modeling task. The final sequence contained $244$ \emph{stable opposed-preference hands}, partitioned into a \emph{context pool of} $97$ \emph{hands} and a \emph{test pool of} $147$ \emph{hands} ($40/60$ split). Within each hand, white-card order was randomized to reduce the influence of known positional tendencies: GPT-4o exhibited a recency bias, while Opus-4.5 exhibited a primacy bias.

The Czar model was \emph{GPT-4o}, and the Player model was \emph{Claude Opus-4.5}. The same 147 test items were evaluated across all five conditions. Conditions 4 and 5 used the 97-item context pool; Conditions 1–3 used no context examples.

\subsection{Overall Accuracy}

\begin{figure*}
\centering
\includegraphics[width=0.6 \textwidth]{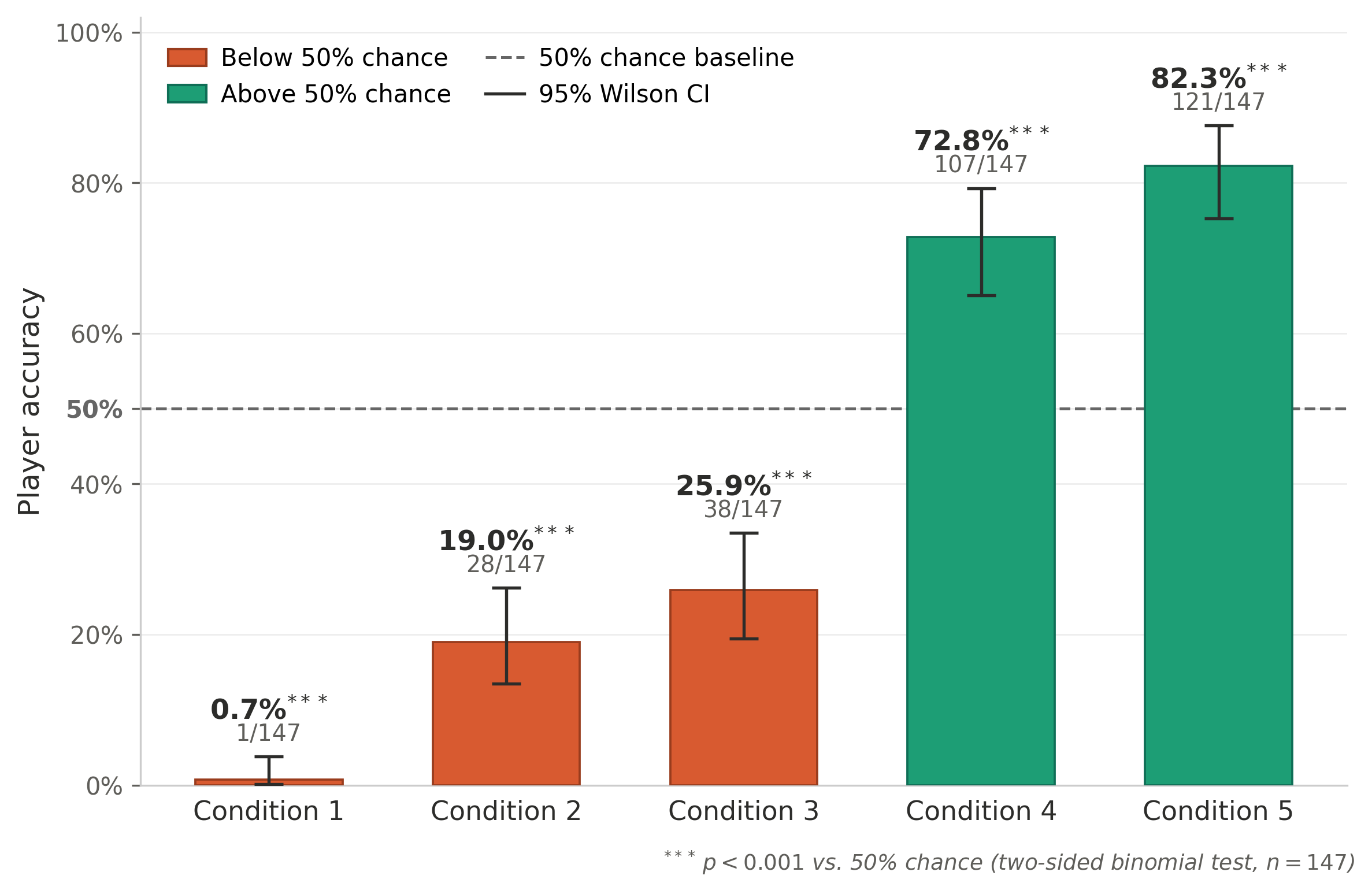}
\caption{Overall Accuracy}\label{fig:overall-accuracy}
\end{figure*}

Player performance increased substantially as more information about the Czar was provided. Each condition is summarized Figure \ref{fig:overall-accuracy} with its observed accuracy, 95\% Wilson score confidence interval, and the p-value from an exact two-sided binomial test against the 50\% chance baseline.

 The Wilson score interval is preferred over the standard normal-approximation ("plus-or-minus") interval because it remains well-behaved at small samples and at extreme proportions — including the near-zero observed value for Condition 1, where the normal approximation would fail.

The progression separates the conditions (discussed in Section \ref{section:experimental-conditions}) into two regimes. Conditions 1–3 remained well below 50\% chance, with generic instructions and model identity producing only modest gains over the default-preference baseline. Conditions 4–5 rose far above chance: direct examples of the Czar's prior behavior produced a large shift, and adding rationales produced the strongest result. All five conditions deviate significantly from chance — Conditions 1–3 significantly below and Conditions 4–5 significantly above — and in each case the 95\% Wilson interval excludes the 50\% chance line entirely.

The near-zero baseline (Condition 1) is the expected validation of the opposed-preference design: by construction, the Player's default preference conflicts with the Czar's stable preference, so any improvement above this baseline must reflect movement away from self-preference. 

\subsection{Statistical Comparisons}

Two complementary forms of statistical analysis were applied to the data: per-condition tests against the 50\% chance baseline (reported in Figure \ref{fig:overall-accuracy}) and repeated-measures tests comparing conditions to each other on the same set of test items.

Because the same 147 test hands were evaluated across all five conditions, the analysis used repeated-measures tests. An omnibus Cochran's Q test showed a significant difference in accuracy across the five conditions, $Q = 294.18$, $\emph{df} = 4$, $p = 1.95 \times 10^{-62}$.

Pairwise McNemar exact tests on the matched test items supported the stepwise pattern observed in the raw accuracies. Each adjacent comparison was significant: Condition 2 outperformed Condition 1, Condition 3 outperformed Condition 2, Condition 4 outperformed Condition 3, and Condition 5 outperformed Condition 4. Figure \ref{fig:selected-mcnemar} shows analysis results for key comparisons.

The largest pairwise shift occurred between Condition 3 and Condition 4. In the McNemar comparison, the Player was correct in Condition 4 but incorrect in Condition 3 on $81$ hands, while the reverse occurred on only $12$ hands. The shift between Condition 4 and Condition 5 was smaller but still significant: the Player was correct in both conditions on $97$ hands, correct in Condition 5 but incorrect in Condition 4 on $24$ hands, and correct in Condition 4 but incorrect in Condition 5 on $10$ hands.

The directional pattern of the per-condition binomial tests, summarized in Table \ref{table:mcnemar-summary}, is consistent with the opposed-preference design: Conditions 1–3 fall significantly below chance, reflecting the limited Czar information available to the Player, while Conditions 4 and 5 rise significantly above chance, indicating that the Player was able to use past Czar behavior — especially when accompanied by rationales — to select the Czar-preferred card on new hands.


\begin{figure*}[htb!]
\centering 

\begin{subfigure}{0.45\textwidth}
\centering
\begin{tabular}{l|ll||l}
                       & C2 \ding{51} & C2 \ding{55} & Total \\ \hline
C1 \ding{51}     &   0           &      1                &    1  \\ \hline
C1 \ding{55}    &   28           &     118           &    146 \\ \hline \hline
Total                &   28           &  119               &    147 \\ \hline
\end{tabular}
\caption{Condition 1 (C1) vs Condition 2 (C2)}
\end{subfigure}
\hfill
\begin{subfigure}{0.45\textwidth}
\centering
\begin{tabular}{l|ll||l}
                       & C3 \ding{51} & C3 \ding{55} & Total \\ \hline
C2 \ding{51}    &   24              &   4                &    28  \\ \hline
C2 \ding{55}    &   14              &  105             &    119 \\ \hline \hline
Total               &   38              &  109             &    147 \\ \hline
\end{tabular}
\caption{Condition 2 (C2) vs Condition 3 (C3)}
\end{subfigure}

\bigskip

\begin{subfigure}{0.45\textwidth}
\centering
\begin{tabular}{l|ll||l}
                       & C4 \ding{51} & C4 \ding{55} & Total \\ \hline
C3 \ding{51}    &   26              &    12             &  38     \\ \hline
C3 \ding{55}    &   81              &    28             &  109   \\ \hline \hline
Total                &  107             &     40          &    147 \\ \hline
\end{tabular}
\caption{Condition 3 (C3) vs Condition 4 (C4)}
\end{subfigure}
\hfill
\begin{subfigure}{0.45\textwidth}
\centering
\begin{tabular}{l|ll||l}
                       & C5 \ding{51} & C5 \ding{55} & Total \\ \hline
C4 \ding{51}     &  97              &   10              & 107     \\ \hline
C4 \ding{55}    &   24              &   16              &  40   \\ \hline \hline
Total               &  121             &   26              & 147   \\ \hline
\end{tabular}
\caption{Condition 4 (C4) vs Condition 5 (C5)}
\end{subfigure}

\caption{Selected McNemar Pairwise Comparison Results}\label{fig:selected-mcnemar}
\end{figure*}

\begin{table*}
\centering
\begin{tabular}{cccccll}
Comparison & Discordant pairs & Favoring earlier & Favoring later & McNemar p    & Sig & Direction \\ \hline
C1 vs C2      & $29$                      &  $1$                      & $28$                   & $< 0.001$       & ***  & C2 outperforms \\ \hline
C2 vs C3      & $18$                      &  $4$                      & $14$                   & $=0.031$        &  *   & C3 outperforms \\ \hline
C3 vs C4      & $93$                      & $12$                     & $81$                   & $< 0.001$       & ***  & C4 outperforms \\ \hline  
C4 vs C5      & $34$                      & $10$                     & $24$                   & $=0.024$        & *     & C5 outperforms \\ \hline
\end{tabular}
\medskip
\caption{McNemar Summary}\label{table:mcnemar-summary}
\end{table*}

\subsection{Summary of Findings}

Three findings emerge. First, the opposed-preference hand set successfully created a strong conflict between the Player's default preference and the Czar's preference: under default preference, the Player almost never selected the Czar-preferred card. Second, generic Czar-modeling instructions and naming the Czar as GPT-4o produced measurable but limited improvement, leaving accuracy below chance. Third, direct evidence of prior Czar behavior produced a large improvement, and adding rationales produced a further gain — suggesting that explanations expose features of the Czar's evaluative pattern beyond the selected cards alone.

Together, these results provide evidence that an LLM acting as Player can move beyond its own default preference and approximate the humor preferences of another LLM, but primarily when given direct examples of the Czar's prior behavior.

\section{Discussion}

Two features of these findings deserve closer interpretation: the limited gains from generic role information, and the large effect of direct behavioral evidence about the Czar.

From a Theory-of-Mind perspective, the experiment captures a simplified version of audience-specific humor prediction. In human CaH play, a skilled Player often brackets their own sense of humor and asks what the current Czar is likely to find funny. The same structure appears here in operational form: the Player must predict the Czar's evaluative perspective rather than express their own. Generic Czar-modeling instructions and model identity produced measurable but limited gains. The Player could partially shift away from self-preference, but lacked enough information to model the Czar accurately.

The strongest evidence for theory-of-mind-like preference modeling appeared when the Player was given examples of GPT-4o's prior evaluations, and stronger still when those examples were accompanied by rationales. A bare selection tells the Player which card the Czar chose; a rationale tells it why, naming the comedic features the Czar treated as salient. Articulated criteria are more transferable than a pattern of selections — they can be applied within a new black card whose comedic possibilities differ from those in the context examples.

The Condition 4-to-5 gain warrants closer examination because the Player received identical sets of past Czar selections in both conditions. The 97 context examples were the same; Condition 5 added GPT-4o's articulated reasoning for each selection. Accuracy rose from 72.8\% to 82.3\% on the held-out test pool of 147 hands, with the McNemar comparison showing 24 hands flipping favorably and 10 unfavorably. Whatever produced this gain must be attributable to something the rationales added beyond what the choices alone provided.

This is informative because the choices themselves are a rich signal. By Condition 4, the Player already had 97 instances of GPT-4o pairing specific black cards with specific selected white cards — substantial behavioral data from which to extrapolate. That this data could be improved upon by accompanying explanations suggests that the rationales are doing more than supplying redundant or marginally useful surface information. They are exposing features of GPT-4o's evaluative pattern that the selections alone do not make visible: that the Czar favored absurdity in one kind of hand, specificity in another, social incongruity in a third. Articulated features can be applied to new hands even when those hands differ from the context examples in their black card, their white card content, or their comedic structure — that is, when they fall within different comedy footprints than those in the context pool. Bare selections, by contrast, transfer best when new hands resemble familiar ones along these dimensions.

Interpreted this way, the Condition 4-to-5 gain is consistent with the Player extracting transferable evaluative criteria from the Czar's articulated reasoning, rather than merely correlating surface features of past choices with new ones. This connects directly to theory of mind. The capacity to extract and apply another agent's evaluative criteria — to grasp what that agent treats as funny, salient, or worth selecting, and to deploy that grasp in new situations — is a recognizable instance of what theory of mind has been concerned with: modeling another mind well enough to predict its judgments. The version of theory of mind at stake here is evaluative rather than epistemic. The Player is not tracking GPT-4o's beliefs about an external situation but its preferences in a domain where preference itself constitutes the answer. Whether the Player's behavior is best characterized as a form of preference modeling, criterion extraction, or sophisticated pattern abstraction is an interpretive question the present experiment does not settle. The behavioral result, however, holds regardless of which characterization one adopts: articulated reasoning from one model provides usable information to another beyond what is contained in the choices themselves, and that information supports a form of cross-model evaluative theory of mind.

These findings should be interpreted cautiously. The experiment does not show that Opus-4.5 possesses human-like Theory of Mind or that it represents GPT-4o as having beliefs, desires, or subjective experiences. The safer interpretation is behavioral: Opus-4.5 used examples and rationales to infer a usable approximation of GPT-4o's preference structure. This is theory-of-mind-like in the sense that the Player's behavior shifted from self-preference to other-preference prediction, but it remains an operational form of preference modeling rather than evidence of human-like social cognition.

Overall, the results suggest that LLMs can perform cross-model audience modeling in a controlled subjective-preference task. Model identity alone was not enough to support strong performance, but direct behavioral evidence substantially improved the Player's ability to predict the Czar's choices. The CaH setting is useful because it makes the distinction between "what I prefer" and "what another evaluator prefers" experimentally visible. In that sense, this experiment does not resolve the broader question of machine Theory of Mind, but it provides a concrete way to study one of its behavioral shadows: the ability to move beyond self-preference and approximate another agent's preferences.

\section{Future Work}

The behavioral evidence the present experiment provides leaves open a number of conceptual and empirical questions about what is being modeled and how. Among the directions worth pursuing are stronger empirical tests of preference modeling: whether a Player's predictions adapt as it observes more of a Czar's choices over time, whether the Player discriminates between Czars when the same cards are held constant across them, whether modeling transfers to surface-dissimilar cases beyond the comedy footprints of the context examples, and whether the Player exhibits counterfactual sensitivity — predicting how a Czar would have responded under altered conditions or why a different card would have lost. These are illustrative rather than exhaustive, but they mark dimensions along which behavioral evidence for modeling, as opposed to surface pattern-matching, could be substantially strengthened.

The most direct next experiment is a multi-Czar extension. Running the same condition gradient against several Czars — different LLMs, or the same LLM under different prompted personas — would test whether the Player's predictions track Czar identity rather than producing globally well-formed humor judgments. This addresses cross-agent discrimination, which the present single-Czar design does not engage.

Beyond stronger empirical tests, the conceptual question of what evaluative theory of mind amounts to in artificial systems — how it relates to the epistemic theory of mind tested in existing benchmarks, and what cognitive notion of "modeling" it requires — remains open. Extension to other subjective-preference domains, including aesthetic and moral judgment, would broaden the empirical base on which that question can be addressed.

\section{Conclusion}

The opposed-preference design used in this paper made cross-model preference modeling experimentally tractable. By isolating hands on which Opus-4.5 and GPT-4o held stable but opposite preferences, the experiment created a setting in which a Player choosing according to its own preference would almost certainly fail. The four remaining conditions could therefore be interpreted as graded tests of whether the Player could move away from that default.

Generic instruction and model identity produced only limited improvement, while direct behavioral evidence about the Czar produced a large increase in performance, and rationales added a further gain. The findings should not be interpreted as evidence of human-like Theory of Mind; a more cautious reading is that the model exhibited theory-of-mind-like behavior in an operational sense, adjusting its selections away from self-preference and toward another agent's demonstrated preferences.

More broadly, the CaH setting highlights a distinction between objective correctness and audience-specific prediction. In tasks involving humor, taste, judgment, or preference, success may depend not on finding the universally best answer but on modeling the evaluator. This paper shows that LLMs can exhibit meaningful movement in that direction when given appropriate evidence, making Czar-preference modeling a promising framework for studying preference generalization, audience modeling, and the behavioral shadows of Theory of Mind in large language models.

\printbibliography

\end{document}